\documentclass[fleqn,twoside]{article}%
\usepackage{amsmath}
\usepackage{amsfonts}
\usepackage{amssymb}
\usepackage{graphicx}
\usepackage{cite}
\def\be{\begin{equation}}
\def\ee{\end{equation}}
\def\bi{\bibitem}
\begin{document}
\title{Bianchi $VI_0$ viscous fluid cosmology with magnetic field.}
\author{Marcelo Byrro Ribeiro$^1$, and Abhik Kumar Sanyal$^2$}
\maketitle
\noindent
\begin{center}
\noindent
$^1$ Observat6rio Nacional, CEP 20921, Rio de Janeiro, Brazil.\\
$^2$Department of Physics, Jadavpur University, Calcutta 700 032, India.\\
\end{center}
\footnotetext{Electronic address:\\
$^2$ sanyal\_ ak@yahoo.com;\\
$^2$ Present address: Dept. of Physics, Jangipur College, India - 742213.}
\noindent
\abstract{A spatially homogeneous Bianchi type $VI_0$ model containing a viscous fluid in the presence of an axial magnetic field has been studied. A barotropic equation of state together with a pair of linear relations among the square root of matter density, shear scalar, and expansion scalar have been assumed. Solutions are obtained in the presence of a magnetic field, only in two special cases, which are comparatively simpler. The complete solutions for this model in the absence of a magnetic field are also obtained. The presence of a magnetic field in the former case however, does not in effect cause any major modification in the fundamental nature of the initial singularity of the expanding model.}

\maketitle
\flushbottom
\section{Introduction:} The investigation of cosmological models in Einstein's theory is usually made by choosing the energy momentum tensor of matter as that due to a perfect fluid. These models lead to an initial singular state. Of course, it is important to investigate more realistic models that take into account dissipative processes due to viscosity.\\

The first suggestion appeared from the investigated by Misner \cite{1}, who proposed that the neutrino viscosity acting in the early era might have considerably reduced the present anisotropy of the black-body radiation (CMBR) during the process of evolution. Murphy \cite{2} in 1973 showed that the bulk viscosity can push the initial singularity in Friedman universe to the infinite past but at the cost of violating the Hawking-Penrose energy conditions. Belinskii and Khalatnikov \cite{3} studied the behavior of anisotropic spatially homogeneous models with viscous fluid in the asymptotic limits. They assumed that the fluid viscosity coefficients could be expressed as power functions of the matter density. They found that the dissipative mechanism due to the presence of viscosity not only modifies the nature of the initial big bang singularity, but also can account for the anomalously large entropy per baryon in the present day universe. Similar properties were shown later by Banerjee, Duttachoudhury, and Sanyal \cite{4} by constructing
particular Bianchi-I model consisting of a viscous fluid. Other models with viscosity terms included in the stress energy tensor were constructed earlier by Banerjee and Santos \cite{5,6}, Banerjee, Duttachoudhury, and Sanyal \cite{7}, Coley and Tupper \cite{8,9} and Santos, Dias, and Banerjee \cite{10}. Also problems with axial magnetic fields in Bianchi-I, Bianchi-III and Kantowski-Sachs viscous fluid models were previously investigated by Banerjee and Sanyal \cite{11}. Though the recently developed theory of inflationary cosmology \cite{12} using GUT, claims to have given a plausible explanation for the outstanding cosmological problems, such as the high degree of isotropy and large entropy per baryon in the present universe, the theory itself appears to be incomplete yet in many aspects. It is therefore worthwhile to investigate if the classical relativity theory is successful in dealing with the above problems by introducing dissipative phenomena in the matter content of the universe.\\

In this paper we proceed to investigate the Bianchi $VI_0$ model filled with a viscous fluid characterized by both bulk and shear viscosities associated with a magnetic field in the axial direction. Evidently the task of obtaining exact solutions in a viscous fluid model becomes more difficult than the corresponding perfect fluid case, due to a larger number of unknown physical quantities to be determined. For this reason, in the present paper we attempt to find exact solutions under the assumption that both the ratios of the shear to expansion rate (${\sigma\over \theta}$) and the density to the square of the expansion (${\rho\over \theta^2}$) are constants. The perfect fluid solutions for the Bianchi-II model with these assumptions were first obtained by Collins and Stewart \cite{13}.\\

In Section 2, we consider Einstein's field equations for a Bianchi $VI_0$ cosmological model and show the dynamical importance of matter density and shear scalar. The entropy variation is also explicitly stated. In Section 3, we obtain two particular solutions in the presence of the magnetic field and complete solutions in the absence of it. Finally we conclude in section 4.

\section{Einstein' field equations and some general results:}

The metric for the spatially homogeneous Bianchi $VI_0$ space time is taken in the following form:
\be\label{2.1} ds^2 = -dt^2 + e^{2\alpha(t)}dx^2 + e^{2\beta(t) +mx} dy^2 + e^{2\gamma(t) - mx} dz^2,\ee
where, $\alpha,~\beta,~\gamma$ are functions of time alone and $m$ is a
constant. The energy momentum tensor for a viscous fluid is

\be\label{2.2} T_{\mu\nu} = (\rho + \bar p)v_\mu v_\nu + \bar p g_{\mu\nu} -\eta U_{\mu\nu}, \ee
\be\label{2.3}\begin{split}&\bar p = p - \left(\zeta - {2\over 3}\eta\right){v^\alpha}_{;\alpha},~~~\mathrm{and}\\&
U_{\mu\nu} = v_{\mu;\nu} + v_{\nu;\mu} + v_{\mu}v^\beta v_{\nu;\beta} + v_{\nu}v^\beta v_{\mu;\beta} \end{split}.\ee
In the above $p$ is the thermodynamic pressure and $\eta$ and $\zeta$ are
the shear viscosity and bulk viscosity coefficients, respectively.
Here $v^\mu$ is the four-velocity vector so that $v_\mu v^\mu = -1$.
Since there is a magnetic field along the $x$ direction, we have
$F_{23}$ as the only non-vanishing component of the electromagnetic
field tensor. From Maxwell's equation it can easily be seen that $F_{23} = A$, where $A$ is a constant of integration. If we have for the stress energy tensor of electromagnetic field the expression
\be E_{\mu\nu} = {1\over 4\pi}\left({F_{\mu}}^{\alpha} F_{\nu\alpha} - {1\over 4}g_{\mu\nu}F_{\alpha\beta}F^{\alpha\beta}\right),\ee
the nonvanishing components are
\be\label{2.4} {E_0}^0 = {E_1}^1= -{E_2}^2 = -{E_3}^3 = -\left(A^2\over 8\pi\right)e^{-2(\beta + \gamma)}.\ee
in the comoving coordinates, $v^\mu = {\delta_0}^\mu$, and under the choice of the unit $8\pi G = c = 1$, Einstein's field equations are now:
\be\label{2.5} {R_\mu}^\nu - {1\over 2} {\delta_\mu}^\nu R = - \left({T_\mu}^\nu + {E_\mu}^\nu\right), \ee
Thus in view of Eq. \eqref{2.5} and using Eqs. \eqref{2.1} - \eqref{2.4}, we find the following equations:

\be\label{2.6a} {9\over 2}\left({\dot R^2\over R^2}\right) - {1\over 2}\left(\dot \alpha^2 + \dot\beta^2 + \dot\gamma^2\right) - m^2 e^{-2\alpha} = \rho + \left({A^2\over 8\pi}\right) e^{-2(\beta + \gamma)},\ee
\be\label{2.6b} \ddot\beta + \ddot\gamma + {3\over 2}\left({\dot R\over R}\right)(\dot\beta + \dot\gamma - \dot \alpha) + {1\over 2}\left(\dot \alpha^2 + \dot\beta^2 + \dot\gamma^2\right)- m^2 e^{-2\alpha} = -(\bar p - 2\eta\dot\alpha) +\left({A^2\over 8\pi}\right) e^{-2(\beta + \gamma)},\ee
\be\label{2.6c}  \ddot\gamma +\ddot\alpha  + {3\over 2}\left({\dot R\over R}\right)(\dot\gamma+ \dot\alpha- \dot \beta) + {1\over 2}\left(\dot \alpha^2 + \dot\beta^2 + \gamma^2\right) - m^2 e^{-2\alpha} = -(\bar p - 2\eta\dot\beta) - \left({A^2\over 8\pi}\right) e^{-2(\beta+ \gamma)},\ee
\be\label{2.6d} \ddot\alpha + \ddot\beta + {3\over 2}\left({\dot R\over R}\right)(\dot\alpha + \dot \beta -\dot\gamma) + {1\over 2}\left(\dot \alpha^2 + \dot\beta^2 + \gamma^2\right) - m^2 e^{-2\alpha} = -(\bar p - 2\eta\dot\gamma) - \left({A^2\over 8\pi}\right) e^{-2(\beta+ \gamma)},\ee
along with $\dot\beta -\dot\gamma = 0$. A dot represents time differentiation and $R$ stands for

\be\label{2.7} R^3 = \exp(\alpha + \beta + \gamma). \ee
In view of Eq. \eqref{2.6a} and with a suitable coordinate transformation
we can have
\be\label{2.8} \beta = \gamma.\ee
Combining the field equations \eqref{2.6a} - \eqref{2.6d} and using Eq.
\eqref{2.8}, we get the following set of equations:

\be\label{2.9a}{1\over 3}\theta^2 - \rho - \sigma^2  = m^2 e^{-2\alpha} + n^2  e^{-4\beta},\ee
\be\label{2.9b} \ddot\beta + (\theta + 2\eta)\dot\beta -{1\over 2}\zeta\theta -{2\over 3}\eta\theta -{1\over 2}(\rho - p)= n^2 e^{-4\beta},\ee
\be\label{2.9c} \ddot\beta + (\theta + 2\eta)\dot\beta + \zeta\theta -{2\over 3}\eta\theta -{1\over \theta^2} - 2\sigma^2 -\dot\theta-(\rho + p)= 2n^2 e^{-4\beta}\ee
Here we have used the relation \eqref{2.9a} to derive the other two. In the set of equations \eqref{2.9a} - \eqref{2.9c}, ${A^2\over 8\pi}$ has been
replaced by $n^2$. The expansion and the shear scalars $\theta$ and $\sigma^2$ are defined in the usual way:

\be\label{2.10} \theta = {v^\alpha}_{;\alpha} = \dot\alpha + 2\dot\beta = 3\left(\dot R\over R\right),\ee
and
\be\label{2.11} 2\sigma^2 = \sigma_{\mu\nu}\sigma^{\mu\nu} = \dot \alpha^2 +2\dot\beta^2 -{1\over 3}\theta^2,\ee
where the shear tensor $\sigma_{\mu\nu}$ has the usual expression
\be\label{2.11a} \sigma_{\mu\nu} = {1\over 2}(v_{\mu;\nu}+ v_{\nu;\mu}) + {1\over 2}\left(v_\mu v^\beta v_{\nu;\beta} +v_\nu v^\beta v_{\mu;\beta}\right) +{1\over 3}\big(g_{\mu\nu} + v_{\mu}v_{\nu}\big)\theta.\ee
From Eqs. \eqref{2.9b} and \eqref{2.9c} we get
\be\label{2.12} \dot\theta = - 2\sigma^2 -{1\over 3}\theta^2 -{1\over 2}\left[\rho + 3(p - \zeta\theta)\right] - n^2 e^{-4\beta},\ee
and the divergence relation $(T^{\mu\nu} + E^{\mu\nu})_{;\nu} = 0$ yields
\be\label{2.13} \dot \rho = -(\rho + p)\theta + \zeta \theta^2 + 4\eta \sigma^2.\ee
It is interesting to observe from Eq. \eqref{2.12} that for a contracting model (that is, for $\theta < 0$) the time derivative for the expansion scalar $\theta$ is less than zero ($\dot \theta < 0$). It means that $\theta$ remains negative always and thus collapse cannot be halted for a physically reasonable fluid ($\rho > 0, p > 0$). On the other hand, if the bulk viscosity $\zeta$ is very small and can be ignored, one has $\dot\theta < 0$ independent of whether the model is expanding or contracting. Thus there may be a maximum but no minimum of the volume. One can easily verify that in both the cases $R_{\mu\nu} v^\mu v^\nu < 0$, and thus Hawking's energy condition is satisfied.\\

Another relation showing the dynamical importance of matter density and shear scalars can be derived in view of the field equations \eqref{2.9a} - \eqref{2.9c} and also using Eqs. \eqref{2.12} and \eqref{2.13}. This expression is explicitly given in the form
\be\label{2.14}\begin{split} \left({\rho\over \theta^2}\right)^\centerdot & =-\left[\left({1\over \theta^2}\right)\left(\sigma^2 + m^2 e^{-2\alpha} + n^2 e^{-4\beta}\right)\right]\\&
=\left({\sigma^2\over \theta^2}\right)\left[3(\rho-p)\theta^{-1} + 3\zeta+4\eta\right] \\&
+\left({m^2 e^{-2\alpha}\over \theta^2}\right)[3\zeta -(\rho + 3p)\theta^{-1}]+\left({n^2 e^{-4\beta}\over \theta^2}\right)[3\zeta + (\rho - 3p)\theta^{-1}].
\end{split}\ee
From the definitions of $\theta$ and $\sigma$ given by Eqs. \eqref{2.10} and
\eqref{2.11} it is possible to write

\be\label{2.15} \sigma^2 = {1\over 3}\theta^2 - \dot\beta(2\theta - 3\dot\beta),\ee
which in turn yields
\be\label{2.16}  \dot \beta = \left({\theta\over 3} \pm {\sigma\over \sqrt 3}\right).\ee
Now differentiating Eq. \eqref{2.15} with respect to time and substituting
$\ddot \beta$ from the field equation \eqref{2.9b} and $\dot\beta$ and $\dot\theta$ from Eqs. \eqref{2.16} and \eqref{2.12}, respectively, one can finally obtain, after a little manipulation and utilizing \eqref{2.9a}, the relation for shear dissipation in the form
\be\label{2.17} (\sigma^2)^\centerdot = -2(2\eta+\theta)^2\pm{4\sigma\over \sqrt 3}\big(n^2 e^{-4\beta} - m^2 e^{-2\alpha}\big).\ee
Using Eq. \eqref{2.16} in the above relation it is possible to obtain
further a very similar kind of relation
\be\label{2.18} {\big(\sigma^2 R^6\big)^\centerdot\over R^6} = -4\eta\sigma^2 - n^2 {\big(e^{-4\beta} R^4\big)^\centerdot\over R^4} - m^2 {\big(e^{-2\alpha} R^2\big)^\centerdot\over R^2}.\ee
The relations \eqref{2.17} and \eqref{2.18} are generalizations of the corresponding equations derived for Bianchi-I space-time in a previous communication \cite{4} Further, as considered by Belinskii and Khalatnikov \cite{3}, let the time derivative of the entropy density be

\be {\dot\Sigma\over \Sigma} = {\dot\rho\over (\rho + p)},\ee
where $\Sigma$ is the entropy density. The total entropy can be defined as $s = R^3 \Sigma$, the time derivative of which can be found in view of Eqs. \eqref{2.13} and the above one, as
\be\label{2.19} {\dot s\over s} = {\big(\zeta\theta^2 + 4\eta \sigma^2\big)\over (\rho + p)}.\ee
Now, since $\rho + p > 0$ and $\zeta > 0$, $\eta > 0$, so $\dot s > 0$, which implies that the total entropy will always increase with the change of proper time irrespective of any model (expanding or contracting).

\section{Exact solutions of Einstein's field equations:}

We have obtained a set of three field equations, viz., \eqref{2.9a} - \eqref{2.9c} with six unknown quantities, viz., $\alpha,~\beta,~\rho,~p,~\eta$ and $\zeta$ to be determined. So in order to obtain exact solutions of the field equations we consider three more physically reasonable equations: one is a barotropic equation of state between matter density and thermodynamic pressure and the other two are a pair of linear relations connecting matter density, expansion, and shear scalars. These are:

\be\label{3.1} p = \epsilon\rho, \hspace{0.5 in}\rho = C^2 \theta^2,\hspace{0.5 in}\sigma^2 = D^2 \theta^2,\ee
where $C$ and $D$ are two constant quantities. Hence the field equations \eqref{2.9a} - \eqref{2.9c} can now be written in view of Eqs. \eqref{2.10} - \eqref{2.13} and \eqref{3.1} as

\be\label{3.2a}  \Big({1\over 3} - C^2 - D^2\Big)\theta^2 = m^2 e^{-2\alpha} + n^2 e^{-4\beta},\ee
\be\label{3.2b} \ddot \beta + (\theta + 2\eta)\dot\beta  - \left[{(1-\epsilon)\over 2}\right]C^2 \theta^2 -{1\over 2}\zeta\theta - {2\over 3}\eta\theta = n^2 e^{-4\beta},\ee
\be\label{3.2c} \ddot \beta + (\theta + 2\eta)\dot\beta  - \left[{1\over 3} + (1+\epsilon)C^2 + 2D^2\right]\theta^2 -\zeta\theta - {2\over 3}\eta\theta -\dot\theta= 2n^2 e^{-4\beta}.\ee
In view of Eqs. \eqref{2.16} and \eqref{3.1} we have
\be\label{3.3} \dot\beta = \left({1\over 3} \pm {D\over \sqrt 3}\right)\theta,  \ee
which, when substituted in Eq. \eqref{2.10}, gives us
\be\label{3.4} \dot\alpha =  \left({1\over 3} \mp {2D\over \sqrt 3}\right)\theta. \ee
The above two equations \eqref{3.3} and \eqref{3.4} lead to the relation $\dot\alpha = a\dot \beta$, where the constant $a$ stands for $a = {\big({1\over 3} \mp {2D\over \sqrt 3}\big)\over\big({1\over 3} \pm {D\over \sqrt 3}\big)}$, so that
\be\label{3.5}  e^{-2\alpha} = b e^{-2 a\beta}.\ee
In Eq. \eqref{3.5} $b$ is an integration constant of positive magnitude. Now in view of Eqs. \eqref{3.3} and \eqref{3.2a} we get
\be {\Big({1\over 3} - C^2 - D^2\Big)\over \Big({1\over 3} \pm {D\over \sqrt 3}\Big)^2}  = m^2 b e^{-2 a\beta} + n^2 e^{-4\beta}, \ee
which can be written as
\be\label{3.6} \dot\beta =\big[C_1e^{-2a\beta} + C_2e^{-4\beta}\big]^{1\over 2},\ee
where, $C_1$ and $C_2$ being two constants. It is not difficult to show that
both $C_1$ and $C_2$ are greater than zero. Writing $x$ for $e^{2\beta}$, relation \eqref{3.6} can be integrated to yield
\be\label{3.7} {1\over 2}\int{dx\over \big[C_1 x^{2-a} + C_2\big]^{1\over 2}} = t - t_0.   \ee
The explicit value for $x$ (that is, $e^{2\beta}$) is obtainable upon choosing specific values for $a$. We consider here two special cases: $a = 2$ and $a = 0$. It is evident that the parameter a cannot be unity because then from its definition $D = 0$ or, in other words, the shear vanishes. When $a = 2$, we have
\be\label{3.8} x = e^{2\beta} = 2\big(C_1 + C_2\big)^{1\over 2}(t-t_0) = C_3(t-t_0),\ee
where $C_3$ is a constant and is equal to $2(C_1 + C_2)^{1\over 2}$. Thus
from Eq. \eqref{3.5}
\be\label{3.9}e^{2\alpha} = \Big({1\over b}\Big)e^{4\beta}. \ee
The expansion scalar $\theta = {2\over t-t_0}$ and the proper volume $R^3 = e^{\alpha + 2\beta} = {{C_3}^2\over \sqrt b}(t-t_0)^2$. From the above solutions, one can conclude that for such a model as $t \rightarrow t_0$, $R^3 = 0$ i.e., the proper volume vanishes, the expansion scalar $\theta \rightarrow \infty$, and in consequence $\rho \rightarrow \infty$, $\sigma^2 \rightarrow \infty$. It is a point singularity. The magnetic field $B$ being proportional to $e^{-2\beta}$ also increases to an indefinitely large value at the singularity. On the other hand as $t\rightarrow \infty$ , we have $\theta \rightarrow 0$, $\rho \rightarrow 0$,$\sigma^2 \rightarrow 0$, and $R^3 \rightarrow \infty$. The second case is for $a = 0$, when we have the integral
\be\label{3.10} {1\over 2}\int{dx\over \big(C_1 x^2 + C_2\big)^{1\over 2}} =(t-t_0).\ee
On integration we obtain, since $C_2 > 0$,
\be {1\over 2\sqrt{C_1}}\ln\big[x\sqrt{C_1} + \sqrt{C_1 x^2 + C_2}\big] = t-t_0, \ee
so that the solution for $e^{2\beta}$ is given by
\be\label{3.11} x = e^{2\beta} = {1\over 2\sqrt{C_1}}e^{-2\sqrt{C_1}(t-t_0)}\big[e^{4\sqrt{C_1}(t-t_0)} - C_2\big].\ee
In this case $\alpha$ = constant. Now as $t \rightarrow t_0$ such that

\be e^{4\sqrt{C_1}(t-t_0)} = C_2,\ee
we have $e^{2\beta} \rightarrow 0$, the proper volume $R^3 \rightarrow 0$, $\theta \rightarrow \infty$, so that$\rho\rightarrow \infty$, $\sigma^2 \rightarrow \infty$, and the magnetic field $B \rightarrow \infty$. On the other hand, as $t$ increases, $e^{2\beta}$ increases and approaches an infinitely large magnitude as $t \rightarrow \infty$. In this limit $\theta \rightarrow 2\sqrt {C_1}$, that is a finite magnitude so that the scalars like $\rho$, $\sigma^2$ etc. also remain finite.\\

Now, eliminating both $\zeta$ and $p$ from Eqs. \eqref{2.12} and
\eqref{2.13} and using Eq. \eqref{3.1} the explicit expression for the
shear viscosity coefficient can be obtained. This is given by

\be\label{3.12} \eta = {1\over 2D^2} \left[\left(C^2 - {1\over 3}\right){\dot\theta\over \theta} - {\theta\over 3}\left(2D^2 + {1\over 3} - C^2 -  {n^2 e^{-4\beta}\over 3\theta}\right) \right].\ee
When the metric is known the exact magnitude of $\eta$ can be calculated independently of any equation of state relating density and pressure of the fluid. But the calculations for the bulk viscosity coefficient $\zeta$ involve pressure and therefore one has to know the pressure in order to write the final form of $\zeta$. Let us assume the barotropic equation of state $p = \epsilon \rho$, as considered earlier in Eq. \eqref{3.1}, to be valid for the fluid under consideration. In this case the relation \eqref{2.13} yields

\be\label{3.13} \zeta = {2\over 3}\left[{\dot\theta\over \theta} +\left({1\over 3}+ 2D^2 + {1+\epsilon\over 2}C^2\right)\theta + {n^2 e^{-4\beta}\over \theta}\right].\ee
We now consider a more simple case where there is no magnetic field. Here one can obtain the exact solution for Bianchi type $VI_0$ spatially homogeneous space-time filled with viscous fluid. Now since a magnetic field is absent, we have $n^2 = 0$ and from Eq. \eqref{3.2a} using \eqref{3.4} we obtain the equation
\be\label{3.14} \left[{{1\over 3}-C^2 - D^2 \over \big({1\over 3}\mp {2\over \sqrt 3} D\big)^2}\right]\dot \alpha^2 = m^2 e^{-2\alpha}.\ee
Integrating Eq. \eqref{3.14} and with a suitable time transformation
we get the solutions for $\alpha$ and $\beta$ as
\be\label{3.15} e^{\alpha} = t,\hspace{0.5 in} e^{\beta} = l_0 t^{1\over \alpha},\ee
where $a$ is the same constant as mentioned in Eq. \eqref{3.5} and $l_0$
is another integration constant. The proper volume and expansion scalars are given by
\be\label{3.16}R^3 = e^{\alpha + 2\beta} = l_0 t^{\big(1+{2\over a}\big)},\ee
and
\be\label{3.17} \theta = {1+ {2\over a}\over t},\ee
where, $ 1 + {2\over a} = {1\over 3 }\mp {2D\over \sqrt 3}$. From (3.17) it is evident that $\dot\theta$ is proportional to $\theta^2$ and further in this case $n^2 = 0$, so that the relations \eqref{3.12} and \eqref{3.13} lead us to the conclusion that both the shear and bulk viscosity coefficients are proportional to the expansion scalar $\theta$. This in turn suggests that these viscosity coefficients are linearly proportional to the square root of the matter density, i.e., $\eta = \eta_0 \rho^{1\over 2},~\zeta = \zeta_0 \rho^{1\over 2}$, with $\eta_0$ and $\eta_0$ being constants. Now since $\big({1\over 3}-C^2 - D \big)> 0$, we have $\big({1\over 3} \pm {D\over \sqrt 3}\big) > 0$, but $\big({1\over 3} \mp {2D\over \sqrt 3}\big)$ may be greater than or less than zero depending on the magnitude of $D$. For an expanding case $\theta > 0$, one finds that $\big(1+{2\over a}\big) >0$, that is, $\big({1\over 3} \mp {2D\over \sqrt 3}\big) > 0$. In this case as $t \rightarrow 0$ we have the proper volume $R^3 \rightarrow 0$ and the expansion scalar $\theta \rightarrow \infty$, so that the density $\rho$, shear $\sigma^2$, and the viscosity coefficients $\eta$ and $\zeta$ all
approach infinitely large magnitudes. It represents a point-like singularity, and the model exploding from a singular state asymptotically approaches an infinite expansion stage at $t\rightarrow \infty$. In this limit $\rho$, $\sigma^2$, $\eta$ and $\zeta$ all vanish.\\

There is one particular case $\Big[1-{2D\over \sqrt 3}\Big] < 0$ when the solution is different. Here  $\big(1+{2\over a}\big) < 0$ and so $\theta < 0$. It represents a contracting model. When $t \rightarrow 0$, the model starts from an infinitely large volume $R^3 \rightarrow \infty$, but since $e^\alpha \rightarrow 0$, while $e^\beta \rightarrow \infty$, at this epoch we may say that the model is initially in the form of an infinite disk at the start of contraction. At $t\rightarrow \infty$ we get$R^3 \rightarrow 0$, but now $e^\alpha \rightarrow \infty$, while $e^\beta \rightarrow 0$, so that the singularity is in the form of a line. The peculiarity of this situation is that at this limit of zero volume the expansion scalar $\theta$ becomes vanishingly small, so that the density shear and the viscosity coefficients all vanish in this limit.

\section{Conclusion:}

In the present paper we analyzed a Bianchi-$VI_0$ model with viscous fluid and in the presence of magnetic field in the axial direction. The viscous fluid is characterized by bulk and shear viscosities. We assumed that ${\sigma^2\over \theta^2} = D^2$ is a constant and ${\rho\over \theta^2} = C^2$ is also one. We obtained solutions for only two special cases in the presence of a magnetic field, which, however, do not change the nature of the singularity. In the absence of the magnetic field complete solutions were obtained. Here the viscosity coefficients $\eta$ and $\zeta$ are found to be power functions of the fluid density being proportional to $\rho^{1\over 2}$. In a particular case of the latter the model is a contracting one with a peculiar feature of the density, viscosity coefficients, shear, etc. all approaching negligible values in the limit of zero volume.\\

\noindent
\textbf{Acknowledgements:}
Thanks are due to Dr. A. Banerjee and to the referee for their valuable suggestions. We also wish to thank U.G.C. (India) and CNPq (Brazil) for financial support.

\end{document}